\documentclass[10pt]{article}

\usepackage[margin=1in]{geometry}
\usepackage{amsmath}
\usepackage{amssymb}
\usepackage{booktabs}
\usepackage{graphicx}
\usepackage{enumitem}
\usepackage{hyperref}
\usepackage[numbers,sort&compress]{natbib}
\usepackage{xcolor}

\hypersetup{
  colorlinks=true,
  linkcolor=blue,
  citecolor=blue,
  urlcolor=blue
}

\newcommand{\system}{\textsc{Floor First}}

\title{Think Before You Grid-Search: Floor-First Triage for LLM Serving}
\author{Yihua Liu\\[2pt] \normalsize Taikang Insurance Group\\ \normalsize\texttt{liuyihua1994@gmail.com}}
\date{July 2026}

\begin{document}
\maketitle

\begin{abstract}
Production teams optimizing large language model (LLM) serving face a configuration space spanning parallelism layout, batching, quantization, sparse attention, and kernel-level work.
The common operational response is to benchmark many configurations and open heavy profilers whenever a latency target is missed.
This paper argues for a different workflow: build an analytical floor first, reconcile benchmarks against that floor, and escalate to profiling only when the residual justifies it.
\emph{Estimation is the analytical layer of profiling}: without it, optimization degenerates to grid search.

We present \system, a residual-driven triage methodology with a zero-dependency artifact: a floor calculator plus an agent skill that makes the discipline enforceable in agentic optimization loops.
The account is compositional---new attention or state-space variants enter by declaring one module, not by rewriting the framework---so the workflow applies to any autoregressive transformer served on accelerators.
\system{} models each decode step as a five-dimensional resource vector (HBM bytes, FLOPs, network bytes, network messages, KV capacity). Terms that use the same resource add; independent resources can overlap. This gives two numbers: an optimistic floor, $\max$, and a no-overlap floor, $\mathrm{sum}$.
Where a measurement falls inside this $[\max,\mathrm{sum}]$ interval is already a diagnostic: it tells us how much overlap the system is getting before any profiler is opened.
Deployment alternatives are then compared by \emph{wall ordering}---which resource wall binds first as load grows---rather than by point benchmarks.

As a case study, we analyze a DeepSeek-V3.2-style 671B MoE/MLA model on 16 NVIDIA H20 GPUs, a hardware point whose ridge point of $\sim$74 FLOP/byte (versus $\sim$590 for H100) makes it an extreme decode-oriented part that no published analysis characterizes.
The floors show that TP16 decoding at batch 64 and 8K context is KV-capacity-limited to $\sim$70 concurrent requests, that DSA-style sparse attention removes the KV-bandwidth term but not the capacity wall, and that an EP16+DP-attention layout trades slightly worse same-batch weight traffic for an order-of-magnitude higher capacity wall ($\sim$644 requests)---while, at cluster-calibrated communication constants, single-stream latency favors TP by $2.4\times$.
The judgment between layouts is thus a computable function of the operating point, which explains why production deployments on identical hardware have shipped opposite attention layouts.
\end{abstract}

\section{Introduction}

In 2025, production teams serving DeepSeek-family models made opposite parallelism choices on comparable hardware.
The official DeepSeek deployment decodes with large-scale expert parallelism and \emph{data-parallel} attention~\citep{deepseek2025profiledata}; vLLM's default DeepSeek recipe now enables the same layout~\citep{vllm2025dprfc}; and a joint LMSYS--Ant Group deployment on the same 16$\times$H20-96G configuration we study reached the same conclusion for decode~\citep{lmsys2025h20}.
Yet TP-sharded attention remains the serving frameworks' historical default, and the production 2$\times$8 H20 cluster whose deployment this paper models serves DeepSeek-V3.2 with plain TP16 attention.
The Ant report states the qualitative rationale---H20 is memory-rich and compute-poor, so decode should exploit bandwidth---but none of these reports derives the choice from an explicit resource account, and none can say at what concurrency the answer flips.
This paper's position is that the disagreement was predictable from a small resource account---roughly five numbers per GPU plus the model dimensions---and that making this account systematic gives a general triage workflow for serving optimization.

The operational problem is broader than one layout choice.
Serving optimization spans model variants, hardware SKUs, tensor parallelism (TP), expert parallelism (EP), data-parallel (DP) attention, prefill/decode disaggregation, continuous batching, sparse attention, quantization, CUDA graphs, communication libraries, and kernels.
Trying combinations on production hardware is expensive, and a benchmark number alone is ambiguous: a slow TPOT can mean a fundamental bandwidth limit, an implementation gap, exposed communication, host overhead, or an underperforming kernel.
Existing research offers increasingly accurate analytical models~\citep{pope2022efficiently,bambhaniya2024genz,davies2025liminal}, simulators and configuration search~\citep{agrawal2024vidur,yao2026kernelsight,xu2026aiconfigurator}, and serving systems~\citep{patel2023splitwise,zhong2024distserve,agrawal2024sarathi}.
What day-to-day performance engineering lacks is a \emph{decision procedure}: when is a measured number close enough to the floor that profiling is wasted effort, and when it is not, where should the profiler look first?

\system{} fills that gap.
The workflow computes per-step floors from model dimensions, calibrated hardware rates, parallelism, and workload shape. It then reads a benchmark as a \emph{residual}: the gap between the measured time and the floor.
If the residual is small, the investigation stops.
If it is large, profiling has a specific job: check whether time is missing between kernels, whether communication is exposed, or whether a kernel class exceeded its analytical budget.
The gap is sharpest for LLM coding agents, whose default optimization behavior is precisely the loop this paper's title warns against; the workflow therefore ships as an enforceable agent skill as well as a human checklist.

\begin{figure*}[t]
  \centering
  \includegraphics[width=0.92\textwidth]{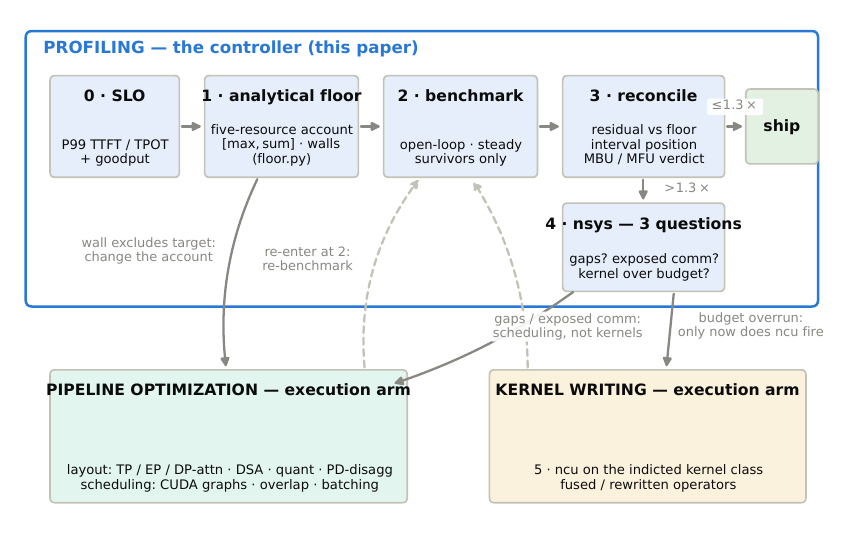}
  \caption{The serving-optimization loop. The floor model decides whether the next action should be layout/scheduling work, kernel work, or no profiling at all. Capacity and wall-ordering verdicts can rule out layouts before GPU time is spent. Nsight Systems is used only when the benchmark leaves a residual that the floor cannot explain; Nsight Compute is reserved for kernel classes that exceed their budget. Every change is re-measured and fed back into the loop.}
  \label{fig:workflow}
\end{figure*}

This draft makes five contributions:

\begin{enumerate}[leftmargin=*]
  \item \textbf{A residual-driven triage workflow} that separates target definition, floor construction, benchmark reconciliation, and profiler escalation, with explicit stopping rules---and whose verdicts dispatch the two execution arms of optimization, pipeline work and kernel work (\S\ref{sec:workflow}, Figure~\ref{fig:workflow}).
  \item \textbf{A two-sided floor model.} Summing contention within a hardware resource and maximizing across resources gives an optimistic floor; the plain sum gives a pessimistic one. The measurement's position inside $[\max,\mathrm{sum}]$ is a zero-cost overlap diagnostic that existing analytical models discard by fixing an overlap assumption (\S\ref{sec:model}).
  \item \textbf{Wall ordering as an output.} Deployment alternatives are compared by which wall---KV capacity, KV bandwidth, weight bandwidth, compute---binds first as load grows, rather than by same-batch latency alone. Capacity is a first-class resource, not a post-hoc memory check (\S\ref{sec:walls}).
  \item \textbf{Metric discipline.} A roofline identity ($\mathrm{MFU}/\mathrm{MBU} = I_{\mathrm{work}}/I_{\mathrm{ridge}}$) dictates which utilization metric is meaningful per phase; we systematize practitioner thresholds for triage verdicts and state their calibration protocol (\S\ref{sec:metrics}, \S\ref{sec:thresholds}).
  \item \textbf{A worked case study and artifact.} We give the first ridge-point characterization of the export-grade H20 GPU and a complete decode account for a DeepSeek-V3.2-style model on 16 H20s, deriving the TP-versus-EP+DP judgment, single-stream physical bounds, and the effect of sparse attention---as executable, zero-dependency Python plus an agent-readable skill (\S\ref{sec:case}).
\end{enumerate}

\paragraph{Status of this draft.}
Case-study numbers are analytical floors computed by the artifact; communication constants are cluster-calibrated (\S\ref{sec:artifact}), GPU rates are datasheet unless marked otherwise.
Section~\ref{sec:experiments-needed} states the measurement campaign required before submission; the residual sweep in particular is planned, not yet performed.
A subsequent revision will fold that sweep back into \S\ref{sec:case}: measured-versus-floor reconciliation plots with per-point $[\max,\mathrm{sum}]$ intervals, the interval-position distribution, and an empirically grounded (or revised) escalation threshold.

\section{Background and Motivation}

\subsection{Serving Metrics and Benchmark Discipline}
\label{sec:benchdiscipline}

LLM serving has at least three operational metrics: time to first token (TTFT), time per output token (TPOT), and goodput---throughput that satisfies a service-level objective (SLO), usually stated on tail latency.
\system{} requires an explicit target before any optimization:
\[
  \text{maximize goodput subject to P99 TTFT and P99 TPOT SLOs}.
\]
Without it, benchmark numbers are not even partially ordered.

Benchmarks that feed the workflow must obey four rules.
(i)~\emph{Open-loop load generation}: closed-loop generators issue the next request only after the previous completes, so an overloaded system automatically receives less load and its tail latency looks healthy. This bias was formalized for classical systems by \citet{schroeder2006open}; LLM serving evaluations rarely revisit the distinction, yet capacity claims are meaningless without it.
(ii)~\emph{Realistic length distributions}, since the prefill:decode ratio and chunked-prefill interference depend on them.
(iii)~\emph{Steady state}: before the KV pool fills, preemption never fires and prefix caches over-hit; the first minutes of a run are honeymoon numbers.
(iv)~\emph{Tail percentiles}, because SLOs live at P99, not the mean.

\subsection{Prefill and Decode Have Different Bottlenecks}
\label{sec:metrics}

Prefill processes many prompt tokens in parallel at high arithmetic intensity and is usually compute-bound; decode streams weights and KV cache per generated token and is usually bandwidth-bound at realistic batch sizes.
This asymmetry underlies phase-splitting systems~\citep{patel2023splitwise,zhong2024distserve,agrawal2024sarathi} and is the basis for metric choice.

Let workload arithmetic intensity be $I_{\mathrm{work}} = \mathrm{FLOPs}/\mathrm{bytes}$ and the hardware ridge point $I_{\mathrm{ridge}} = F_{\mathrm{peak}}/B_{\mathrm{HBM}}$~\citep{williams2009roofline}.
With model FLOPs utilization $\mathrm{MFU} = \mathrm{FLOPs}/(t\,F_{\mathrm{peak}})$ and model bandwidth utilization $\mathrm{MBU} = \mathrm{bytes}/(t\,B_{\mathrm{HBM}})$~\citep{databricks2023mbu},
\[
  \frac{\mathrm{MFU}}{\mathrm{MBU}} = \frac{I_{\mathrm{work}}}{I_{\mathrm{ridge}}}
\]
holds identically.
The two utilizations are not independent; their ratio is fixed by the workload.
For the case study of \S\ref{sec:case}, decode at batch 64 has $I_{\mathrm{work}} \approx 11$ FLOP/byte against an H20 ridge of 74: when MBU reaches 80\%, MFU \emph{equals} 12\%.
Low decode MFU is workload geometry, not a tensor-core bug.
The metric-selection principle is: \emph{report the utilization whose ceiling is 100\% for the binding resource}---MBU for bandwidth-bound decode, MFU for compute-bound prefill.

\subsection{The Workflow: One Controller, Two Execution Arms}
\label{sec:workflow}

Production inference optimization has three arms: \emph{profiling} (knowing where time goes and how far from the limit it is), \emph{kernel writing} (driving a single operator to the hardware limit), and \emph{pipeline optimization} (parallelism layout, batching, overlap, scheduling).
In this paper, profiling is not just ``open Nsight.''
It is the decision loop that decides whether kernel work or pipeline work is worth doing. Kernel work without that decision can optimize an operator that is not on the critical path; pipeline tuning without it becomes grid search.
Figure~\ref{fig:workflow} shows the loop. A capacity or wall-ordering verdict can send us directly to a new layout before we spend GPU time. A residual caused by poor overlap or scheduling sends us to the pipeline arm. Only a kernel class that exceeds its budget sends us to Nsight Compute and kernel work.
After each change, we benchmark again and compare the new result with the floor.

Within the controller, \system{} treats instrumented profiling as exception handling, not as a pipeline stage:

\begin{enumerate}[leftmargin=*]
  \item Define the objective: P99 TTFT, P99 TPOT, goodput.
  \item Compute analytical floors and capacity walls for candidate configurations; discard configurations whose walls exclude the target operating point.
  \item Benchmark the survivors under the discipline of \S\ref{sec:benchdiscipline}.
  \item If the measurement is within the escalation threshold of the floor ($1.3\times$ in our practice), stop---or move to a different optimization axis.
  \item Otherwise open Nsight Systems and answer exactly three questions: are there gaps between kernels (host/scheduling/launch)? is communication exposed rather than overlapped? which kernel class exceeds its analytical budget?
  \item Open Nsight Compute only for the kernel class that the budget table indicts, after gap and overlap causes are excluded.
\end{enumerate}

\subsection{Triage Thresholds and Their Status}
\label{sec:thresholds}

Our operating rules: decode MBU $>70\%$ means the implementation is near the floor on the current axis; 40--70\% suggests overlap or scheduling recovery, worth an Nsight Systems pass; $<40\%$ suggests a system-level defect (host-bound execution, missing CUDA-graph coverage, interference) that per-kernel work will not fix.
For MoE prefill we lower the MFU bands to $50\%/25\%$ because all-to-all communication and expert imbalance are structural.

These numbers systematize practitioner experience~\citep{databricks2023mbu} rather than derive from first principles, and we state their epistemic status plainly: they are \emph{calibration targets}, not constants of nature.
The calibration protocol is to collect (predicted floor, measured, post-hoc root cause) triples across configurations and check two properties: below the threshold, Nsight sessions should yield no actionable finding (no false negatives); above it, a root cause should be identifiable (no wasted escalations).
\S\ref{sec:experiments-needed} lists this as the first required experiment; until then the thresholds should be read as defaults that worked in our deployments, in the same spirit that Top-Down analysis shipped with fixed decision-tree cutoffs.

\section{Analytical Model}
\label{sec:model}

\subsection{Resource Vector and the Two-Sided Floor}
\label{sec:twosided}

Each decode step consumes a five-dimensional resource vector
\[
  r = (\mathrm{HBM\ bytes},\ \mathrm{FLOPs},\ \mathrm{network\ bytes},\ \mathrm{network\ messages},\ \mathrm{KV\ capacity}).
\]
The first four convert to time via calibrated rates; the fifth bounds feasible batch size and enters the goodput judgment (\S\ref{sec:capacity}).
Two aggregation rules follow from the hardware, not from convention:
terms contending for the \emph{same} resource add (weight reads and KV reads share HBM; bandwidth cost and per-message latency share the NIC), while \emph{independent} engines (HBM, SMs, NIC) run concurrently under perfect overlap, so only the slowest shows:
\[
  t^{\mathrm{opt}} = \max(t_{\mathrm{HBM}},\ t_{\mathrm{compute}},\ t_{\mathrm{network}}),
  \qquad
  t^{\mathrm{sum}} = t_{\mathrm{HBM}} + t_{\mathrm{compute}} + t_{\mathrm{network}}.
\]

We keep both numbers because the gap between them is useful.
If a measured step lands near $t^{\mathrm{opt}}$, the hardware engines are already overlapping well and the next gain must come from reducing the dominant account itself. If it lands near $t^{\mathrm{sum}}$, the system is behaving as if the engines were serialized, so scheduling and overlap are the first suspects. If it lands above $t^{\mathrm{sum}}$, overlap cannot explain the measurement; time is leaking into costs outside the model, such as host gaps, preemption, or stragglers.
Existing analytical models collapse this information by choosing an overlap assumption up front (e.g., compute hidden under memory with communication fully exposed~\citep{davies2025liminal}); the interval keeps overlap as something the benchmark can reveal.
One caveat is structural: at batch~1 the layer-serial dependency chain leaves little to overlap, so the honest single-stream floor is $t^{\mathrm{sum}}$, not $t^{\mathrm{opt}}$ (\S\ref{sec:singlestream}).

\subsection{Decode FLOPs}

Per token, parameter GEMMs cost $\approx 2P_{\mathrm{act}}$ FLOPs, where $P_{\mathrm{act}}$ is the activated parameter count ($P_{\mathrm{act}} = P$ for dense models); attention score/value products add a term that depends on context length and attention architecture:
\[
  \mathrm{FLOPs}_{\mathrm{decode}} \approx 2 P_{\mathrm{act}} B + \mathrm{FLOPs}_{\mathrm{attn}}(B, S).
\]
In the decode regimes we study, attention FLOPs matter for the compute row, but the attention path is still dominated by reading KV from HBM. The FLOP account is therefore used mainly to check whether compute can ever become the wall (\S\ref{sec:walls}).

\subsection{Decode Bytes: the MoE Union Correction}
\label{sec:union}

Decode HBM traffic per step is weight traffic plus KV-read traffic.
For MoE models the two accounts diverge: FLOPs follow \emph{activated} parameters, but weight bytes follow the \emph{union} of experts activated across the batch, because a weight tile is read once per step regardless of how many tokens use it.
Under uniform routing with $E$ routed experts and top-$k$ selection, the expected distinct-expert fraction per layer is
\[
  f_{\mathrm{expert}}(B) = 1 - \left(1 - \tfrac{k}{E}\right)^{B},
  \qquad
  f_{\mathrm{expert}}(B) \le \min(1,\ kB/E),
\]
so a batch of 64 touches $\approx 87\%$ of experts for $E{=}256$, $k{=}8$, and the conservative $\min$ bound saturates.
This closed form was previously employed to analyze speculative-decoding gains for MoE~\citep{huang2025moesd} and in-batch expert sharing~\citep{vankov2026xshare}; we apply it to deployment-sizing byte accounts, where it corrects both the weight-bandwidth wall and the compute-wall location by the MoE sparsity ratio (\S\ref{sec:walls}).
Load imbalance makes the uniform-routing expectation optimistic; \S\ref{sec:limits} discusses the failure mode.

KV bytes depend on attention architecture and sharding.
For GQA, the KV cache partitions across up to $n_{\mathrm{kv}}$ heads under TP.
For MLA-style latent attention the latent KV is shared by \emph{all} heads: head-parallel TP cannot shard it, so every TP rank stores and reads the full cache---the observation first stated in engineering channels~\citep{sglang2024v04,vllm2025dprfc} and in recent architecture work~\citep{tang2025tpla,zadouri2025hardwareefficient,meng2026gqla}, whose responses are to redesign the attention or its sharding.
We take the deployment-side view: given unmodified MLA, the replication cost makes DP attention the structurally favored high-concurrency layout, and \S\ref{sec:case} quantifies by how much.

\subsection{Communication}
\label{sec:communication}

TP decode performs two all-reduces per layer; with ring all-reduce the per-op traffic is $2\frac{n-1}{n} B H b_{\mathrm{act}}$ bytes, and per-op time is bandwidth plus a latency term that dominates at small batch:
\[
  t_{\mathrm{AR}} = N_{\mathrm{AR}} \left( \frac{2\frac{n-1}{n} B H b_{\mathrm{act}}}{B_{\mathrm{net}}} + \ell \right).
\]
EP with DP attention replaces all-reduces with dispatch/combine all-to-alls per MoE layer, with traffic $\approx B L_{\mathrm{MoE}}\, r H (b_{\mathrm{disp}} + b_{\mathrm{comb}})$, where $r = n\bigl(1-(1-1/n)^{k}\bigr)$ is the expected number of distinct expert ranks touched by a token's top-$k$ selection ($r \approx 6.5$ at $k{=}8$, $n{=}16$)---the rank-level sibling of the expert-union expectation of \S\ref{sec:union}.
Dispatch sends one activation per touched rank; combine returns one partial sum per touched rank (an upper bound if the implementation aggregates in-network).
The implementation constant is decisive: a low-latency RDMA all-to-all approaches the floor while a naive one can be several times slower.
\system{} therefore separates two questions that the literature usually merges: \emph{which route has the better ceiling}, and \emph{does a near-ceiling implementation of that route exist}?
Comparisons are made at the ceiling level first; only routes that pass are checked for mature implementations (e.g., DeepEP-class all-to-all), and routes must never be eliminated by benchmarking a poor implementation of them.

\subsection{Capacity Wall}
\label{sec:capacity}

The KV capacity wall bounds concurrency:
\[
  B_{\max} = \frac{(C_{\mathrm{HBM}} - C_{\mathrm{weights}} - C_{\mathrm{overhead}})\cdot s_{\mathrm{KV}}}{S\cdot \mathrm{bytes}_{\mathrm{KV/token}}},
\]
where the KV sharding factor $s_{\mathrm{KV}}$ is 1 for MLA under head-parallel TP (full replication), $\min(n, n_{\mathrm{kv}})$ for GQA under TP, and $n$ for DP attention.
Because decode goodput is $B/t_{\mathrm{step}}$ under a TPOT SLO, a layout with slightly worse same-batch $t_{\mathrm{step}}$ but an order-of-magnitude larger $B_{\max}$ can dominate the goodput frontier.
Treating capacity as a fifth resource dimension---rather than an afterthought OOM check---is what turns the account into a deployment judgment.

\subsection{Compositionality: New Architectures Enter by One Module}
\label{sec:compositional}

The account is modular.
Each layer type---attention, routed FFN, dense FFN, or a future state-space block---declares five things: weight bytes, state bytes, per-step state reads, FLOPs, and communication. These quantities are functions of workload $(B,S)$ and of that module's sharding plan. A model is then just a list of modules, and a step account is the per-resource sum over that list.
This design has two consequences.
First, adding an architecture does not require a new framework. DSA changes the attention module's state-read function from $O(S)$ to approximately $O(k)$ per query (\S\ref{sec:walls}); a state-space layer would declare a constant-size recurrent state instead of a growing KV cache, and the capacity wall of \S\ref{sec:capacity} would update automatically.
Second, shardability belongs to the module. MLA's latent KV declares itself unshardable across attention heads, so DP attention follows naturally for high-concurrency serving; architecture work that re-enables TP for latent attention~\citep{tang2025tpla,meng2026gqla} is, in this language, changing that declaration.
Parallelism plans are therefore per-module assignments rather than global labels. Production deployments already live in this space: the LMSYS--Ant system serves prefill with single-node TP8 attention and decode with DP attention plus EP16 on the same hardware~\citep{lmsys2025h20}---a per-phase, per-module plan assignment.
For standard families, the declarations are mostly data entry: the constants come from published model dimensions, as in the artifact's specification file for dense, GQA, MLA, and MoE models. The current artifact hard-codes the two layouts in the case study; the per-module plan interface is staged as the next revision (\S\ref{sec:limits}).

\section{Artifact}
\label{sec:artifact}

The artifact is deliberately small: a specification module (\texttt{specs.py}) holding model, GPU, and cluster constants with datasheet and calibrated rates kept separate; a floor calculator (\texttt{floor.py}) producing the resource table, $[\max,\mathrm{sum}]$ floors, capacity wall, single-stream bound, and prefill floor; a reconciliation tool (\texttt{mbu\_mfu.py}) converting measured TPOT/TTFT into MBU/MFU with a triage verdict; and an agent-readable workflow document (\texttt{SKILL.md}).
The implementation is pure Python with no runtime dependencies; Appendix~A gives the commands used to reproduce the case-study tables.

Calibration is outside the model by design.
The chain is: HBM bandwidth via \texttt{nvbandwidth}; GEMM rates via the production serving shapes; communication via \texttt{nccl-tests} and the production all-to-all's own benchmark.
Datasheet floors prune the design space; calibrated floors make residuals meaningful.

We executed this chain on the target cluster (fallback tools where installation was constrained: a torch-level device copy for HBM, production DeepGEMM kernels for FP8).
GPU-side rates landed near expectations (81\% of datasheet HBM bandwidth under the fallback tool, 89\% of FP8 peak), but calibration moved \emph{every} communication constant: the effective all-reduce rate at decode message sizes is 43 GB/s against a 100 GB/s per-node line rate; small-message latency is $\approx$33 $\mu$s and equal within noise across and within nodes, indicting NCCL launch overhead rather than the wire; and DeepEP's op-level dispatch latency is 60 $\mu$s against a $\sim$30 $\mu$s pure-transport bound.
A raw-NCCL all-to-all on the same fabric measured 12.6 GB/s where DeepEP approaches line rate---a concrete example of the ceiling-versus-implementation distinction in \S\ref{sec:communication}.
One accounting rule emerged: the bandwidth constant must be the latency-\emph{excluded} slope, not any single message size's effective bus bandwidth, or the latency term is double-counted.
Fully calibrated floors are 15--24\% looser than datasheet ones and serve as the reconciliation baseline for \S\ref{sec:experiments-needed}.
Notably, although calibration moved every communication constant---one by $8\times$---no optimistic floor and no deployment judgment of \S\ref{sec:case} changed: on this hardware the HBM account and the capacity wall carry the verdicts, so they are insensitive to precisely the inputs that are hardest to know in advance.
This robustness is a property of wall ordering, not luck: a judgment carried by the first wall survives errors in the accounts of walls never reached.

The \texttt{SKILL.md} component addresses a distinct failure mode: coding agents, like humans, reflexively reach for benchmarks and profilers.
Encoding the workflow as an agent skill makes the floor-first discipline enforceable in agentic optimization loops~\citep{ding2025asap,li2026agentskill}, where the artifact serves as the deterministic arithmetic layer beneath model-driven judgment: a triggered agent must produce the floor account before proposing any benchmark, and profiler use must cite a residual above threshold.
This is not speculative: the calibration campaign of \S\ref{sec:experiments-needed}-(1) was executed on the production cluster by a coding agent operating under this skill---following the measurement protocol, hitting the documented fallback paths when tooling was unavailable, and returning the spec-keyed calibration file this paper consumes.
The artifact, calibration data, and measurement scripts will be released with the next revision of this draft.

\section{Case Study: DeepSeek-V3.2-Style MoE on 16$\times$H20}
\label{sec:case}

\subsection{Hardware Portrait: the H20 Ridge Point}
\label{sec:ridge}

The H20 is the export-compliant Hopper variant: compute is cut to $\sim$15\% of H100 while memory bandwidth is comparable or higher (Table~\ref{tab:ridge}).
The consequence is a ridge point of $296/4.0 \approx 74$ FLOP/byte versus $\approx 590$ for H100: \emph{the H20's memory-bound region is $8\times$ wider}.
A workload below 74 FLOP/byte is memory-bound on H20, so the compute cut is mostly hidden. Above 74, H20 starts to pay for its weaker tensor throughput; beyond the H100 ridge, both GPUs are compute-bound and the full compute gap appears.
Decode at realistic batch sizes sits below the H20 ridge, while prefill sits far above it---so the H20 is naturally suited to decode and weak for prefill.
This single number reframes deployment: deployment reports observe that H20 serves memory-bound decode at flagship-class rates and reason qualitatively from that fact~\citep{lmsys2025h20}, and kernel work targets its weak tensor throughput~\citep{dege2025flashmlaetap}, but to our knowledge no published account derives the deployment consequences from the ridge geometry.

\begin{figure}[t]
  \centering
  \includegraphics[width=0.72\linewidth]{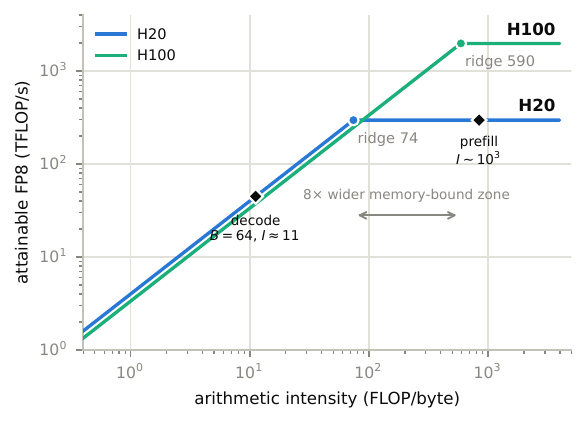}
  \caption{FP8 rooflines of H20 and H100 with the case study's operating points. Decode at $B{=}64$ ($I\!\approx\!11$ for the TP16 account) sits below the H20 ridge, so it is governed by memory traffic rather than tensor peak. Prefill ($I\!\sim\!10^{3}$) sits past both ridges, where the H20 pays its compute cut. The $8\times$ ridge gap is the geometric content of ``a decode part.''}
  \label{fig:roofline}
\end{figure}

\begin{table}[t]
  \centering
  \caption{Datasheet portrait (Figure~\ref{fig:roofline} renders the geometry). The H20 trades compute for nothing else; its ridge point is $8\times$ lower than H100's, widening the memory-bound comfort zone by the same factor.}
  \label{tab:ridge}
  \begin{tabular}{lrrrr}
    \toprule
    & HBM & BW & FP8 dense & Ridge (FLOP/B) \\
    \midrule
    H100 SXM & 80 GB & 3.35 TB/s & 1979 T & $\sim$590 \\
    H800     & 80 GB & 3.35 TB/s & 1979 T & $\sim$590 \\
    H20      & 96 GB & 4.00 TB/s & 296 T  & $\sim$74  \\
    \bottomrule
  \end{tabular}
\end{table}

\subsection{Setup}

We analyze a DeepSeek-V3.2-style model---671B total / 37B activated parameters, FP8 weights, MLA latent KV (70~KB per token across 61 layers), 58 MoE layers with 256 routed experts and top-8 routing---following the public V3-family dimensions~\citep{deepseek2024v3,zhao2025insights} and the DSA sparse-attention direction of V3.2~\citep{deepseek2025v32}.
Deployment: 2 nodes $\times$ 8 H20, 4$\times$200\,Gb HDR InfiniBand per node ($\approx$100 GB/s unidirectional aggregate).
Communication constants are calibrated on this cluster (\S\ref{sec:artifact}); GPU-side rates in Table~\ref{tab:h20-case} stay at datasheet values for reproducibility, with fully calibrated floors 15--24\% looser (\S\ref{sec:artifact}).
The workload is steady-state decode at $B{=}64$, $S{=}8192$.
Table~\ref{tab:h20-case} gives the floors; all values reproduce from the artifact.

\begin{table*}[t]
  \centering
  \caption{Analytical decode floors, DeepSeek-V3.2-style MoE on 16$\times$H20, $B{=}64$, $S{=}8192$ (ms per step; conservative expert union; GPU rates at datasheet, collective rates cluster-calibrated---no meaningful datasheet exists for effective collective bandwidth). The union expectation (\S\ref{sec:union}) tightens weight terms by $\sim$13\% at this batch.}
  \label{tab:h20-case}
  \begin{tabular}{lrrrrrrr}
    \toprule
    Configuration & Weight & KV & HBM $\Sigma$ & Compute & Network & Floor $[\max,\mathrm{sum}]$ & $B_{\max}$ \\
    \midrule
    TP16                    & 10.48 & 9.21 & 19.70 & 2.99 & 8.91 & [19.7,\ 31.6] & 70 \\
    TP16 + DSA-style sparse & 10.48 & 2.30 & 12.79 & 1.50 & 8.91 & [12.8,\ 23.2] & 70 \\
    EP16 + DP attention     & 14.70 & 0.58 & 15.28 & 2.99 & 10.66 & [15.3,\ 28.9] & 644 \\
    \bottomrule
  \end{tabular}
\end{table*}

\subsection{Wall Ordering}
\label{sec:walls}

The account in Table~\ref{tab:h20-case} is best read not row-by-row but as an ordering of the walls a deployment hits as load grows (Table~\ref{tab:walls}).

\begin{table}[t]
  \centering
  \caption{Wall ordering for TP16 at 8K context. Each wall is quantified by the floor account; each removal reshuffles the remaining order. The compute wall is unreachable in the feasible region: the capacity wall caps $B$ at 70, before even the attention-aware compute knee (roughly $B\!\approx\!225$ at 8K, and higher after DSA). The GEMM-only MoE-corrected knee is $B^{*}\!\approx\!670$.}
  \label{tab:walls}
  \begin{tabular}{llll}
    \toprule
    \# & Wall & Binds when & Removed by \\
    \midrule
    1 & KV capacity  & $B \gtrsim 70$ & DP attention ($\to$644) \\
    2 & KV bandwidth & long $S$ (9.2 ms @8K) & DSA ($\to$2.3 ms) \\
    3 & Weight bandwidth & $B \gtrsim 32$ (union sat.) & quantization; EP scale-out \\
    4 & Compute & $B^{*}_{\mathrm{GEMM}} \approx 670$; incl. 8K MLA $\sim$225 & \emph{capacity comes first} \\
    \bottomrule
  \end{tabular}
\end{table}

Two structural facts fall out.
First, the compute wall is not the next wall. If we compare only parameter GEMMs with full-expert weight reads, the dense rule of thumb $2B \ge I_{\mathrm{ridge}}$ would put the knee at $B \approx 37$. For MoE decode, however, each full weight read supports only the activated fraction of the model, so the GEMM-only knee shifts by the sparsity ratio $671/37 \approx 18\times$ to $B^{*}_{\mathrm{GEMM}} \approx 670$.
Including the 8K MLA attention FLOPs lowers the compute-vs-weight knee to roughly $B \approx 225$ (roughly $450$ with DSA), but this still lies beyond the TP16 capacity wall at $B \approx 70$. Once KV bytes are included, the HBM account grows at least as fast as the compute account, so the feasible TP16 region ends before compute can bind.
Chasing decode MFU here is optimizing a wall the deployment cannot reach.
Second, walls are removed in order or not at all: DSA removes the KV-bandwidth wall (row 2) but not the capacity wall (row 1), because top-$k$ selection reduces \emph{reads}, not \emph{residency}---the full KV must remain addressable for future steps.
(The account models DSA's dominant effect, the KV-read reduction. Its lightning indexer still has an $O(S)$ scoring path; we treat it as second-order at 8K but expose it as a module that must be added for long-context accounting~\citep{deepseek2025v32,zhou2026misa}.)
After DSA, the exposed wall is routed-weight bandwidth, which is why quantization and EP-style weight sharding, not more attention work, are the next levers on this hardware.

\subsection{TP16: Both KV Terms Replicate}

TP16 shards all weights 16 ways ($\sim$42 GB/GPU at full expert union) but MLA replicates KV: each GPU reads the \emph{entire} 36.8 GB of active KV per step (\S\ref{sec:union}).
The HBM account is 19.7 ms---weight and KV reads add on the same resource---dominating compute (3.0 ms, of which 9.4 of the 14.2 TFLOP is per-layer MLA attention) and network (8.9 ms: 4.9 ms of traffic at the calibrated 43 GB/s effective all-reduce rate, plus 4.0 ms of latency, $122 \times 33\,\mu$s).
The latency constant is itself a calibration finding: cross-node and intra-node small-message latency measure equal within noise ($\approx$33 vs.\ 32.7 $\mu$s), identifying ungraphed NCCL launch overhead---not wire time---as the per-op cost, with CUDA-graphed production paths reaching $\sim$10 $\mu$s.
Sparse attention cuts the KV term to 2.3 ms, but the capacity wall stays at $B_{\max} \approx 70$ for 8K contexts: DSA changes what is read, not what is stored.

\subsection{EP16 + DP Attention: Trading Bandwidth for Capacity}

EP16 shards routed experts but replicates the non-routed $\sim$18 GB on every GPU, raising per-GPU weight traffic to 58.8 GB (14.7 ms)---worse than TP16's 10.5 ms at equal batch.
DP attention, however, shards both KV \emph{reads} (0.58 ms) and KV \emph{residency}: the capacity wall moves from 70 to $\approx 644$ concurrent 8K requests, an order of magnitude.
Under a TPOT SLO, goodput is $B/t_{\mathrm{step}}$ along the feasible region; the layout with the higher wall wins the frontier even where its same-batch step time is worse.
This is the quantitative content of the deployment folklore ``attention should go DP for MLA serving''~\citep{sglang2024v04,deepseek2025profiledata}: the judgment follows from rows 1--3 of the wall table, needs no benchmark to reach, and---as the next subsection shows---reverses at low concurrency.

\begin{figure}[t]
  \centering
  \includegraphics[width=0.78\linewidth]{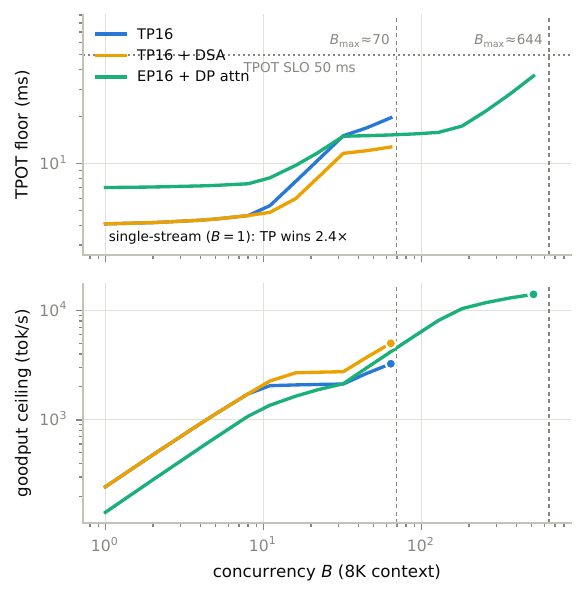}
  \caption{Analytical TPOT floors (top) and goodput ceilings (bottom) versus concurrency at 8K context, computed by the artifact. The KV-capacity walls (dashed) terminate each layout's feasible region: TP16 variants end at $B\!\approx\!70$ while EP16+DP attention continues to $B\!\approx\!644$, dominating the high-concurrency frontier---yet at $B{=}1$ the ordering inverts and TP16 leads by nearly $2\times$. The deployment judgment is a function of the operating point.}
  \label{fig:frontier}
\end{figure}

\subsection{Single-Stream Bounds: the Judgment Inverts}
\label{sec:singlestream}

At $B{=}1$ the layer-serial chain removes overlap opportunity, so the honest floor is the sum (\S\ref{sec:twosided}).
For TP16: weight reads shrink to the touched-expert union ($\sim$2.4 GB, 0.6 ms) but 122 all-reduce latencies cost 4.0 ms at the measured 33 $\mu$s/op---latency, not bandwidth, is the single-stream wall---giving TPOT $\ge 4.9$ ms, i.e.\ $\lesssim 205$ tok/s.
For EP16+DP: at $B{=}1$, the active attention rank reads the replicated non-routed block plus the touched routed experts ($\sim$19.3 GB total, 4.8--4.9 ms), and the step still pays 116 all-to-all latencies at the measured 60 $\mu$s (7.0 ms). The resulting bound is TPOT $\ge 11.8$ ms, or $\lesssim 85$ tok/s---\emph{TP wins the single-stream regime by $2.4\times$}, the mirror image of the high-concurrency judgment.
Adding GPUs helps neither: because the calibration shows the 33 $\mu$s is launch-dominated (\S\ref{sec:case}), the levers are CUDA-graph launch elimination, fewer inter-layer synchronizations, or a faster interconnect---in that order of accessibility.

Two framing notes.
These are physical bounds, not achievable expectations: recent measurements show batch-1 decode on high-bandwidth parts reaching only $\sim$27\% of its analytic floor due to kernel-launch overheads~\citep{chen2026memorybound}, so single-stream floors bound feasibility studies (``can this SLO ever be met?'') rather than predict production numbers.
And the regime dependence of the TP-vs-DP judgment---TP at low concurrency, DP at high---is exactly the shape of the production disagreement in \S1: the cluster we model (TP16 attention) and the official DeepSeek and LMSYS--Ant deployments (DP attention) sit at different operating points on the same frontier.
The account does not say one team erred; it says the choice is a computable function of the target concurrency, and computes it.

\subsection{Reconciliation: Reading Residuals}

% —— v2 时放回：fig_reconcile / fig_interval（等 B 包 residuals.csv，去水印后启用）——
% \begin{figure}[t]
%   \centering
%   \includegraphics[width=0.72\linewidth]{figs/fig_reconcile.pdf}
%   \caption{Reconciliation view for the measurement sweep of \S\ref{sec:experiments-needed}: each configuration's measured TPOT against its predicted optimistic floor, with the $[\max,\mathrm{sum}]$ interval drawn as a vertical band per point. Points inside their band are explained by overlap quality alone; points above the $1.3\times$ line escalate to the profiler.}
%   \label{fig:reconcile}
% \end{figure}
% \begin{figure}[t]
%   \centering
%   \includegraphics[width=0.80\linewidth]{figs/fig_interval.pdf}
%   \caption{The same sweep read as positions inside the $[\max,\mathrm{sum}]$ interval: 0 is perfect engine overlap, 1 is fully serialized, and beyond 1 the step time cannot be explained by overlap at all---time is leaking outside the resource account (host gaps, stragglers, preemption).}
%   \label{fig:interval}
% \end{figure}

Reconciliation compares floors against steady-state \emph{service time} (P50 TPOT), not tail latency: queueing delay is real but lives outside the resource account (\S\ref{sec:limits}), so P99 validates the SLO while P50 diagnoses the engine---conflating them reads tail queueing as kernel residual.

Suppose TP16 (no sparse attention) measures 25 ms P50 TPOT at this workload.
The artifact reports $\mathrm{MBU} = (41.9 + 36.8)\,\mathrm{GB} / (25\,\mathrm{ms} \times 4.0\,\mathrm{TB/s}) = 78.8\%$; the residual is $1.27\times$ the optimistic floor---below the escalation threshold---and the position inside $[19.7, 31.6]$ is $0.45$.
The two signals say different things, which is why both exist: the residual says \emph{stop}---no profiler session is warranted, and the next win must come from changing the account itself (DSA, quantization, or layout)---while the mid-interval position bounds what better overlap could ever recover ($\sim$5 ms here) if that axis were pursued anyway.
Had the measurement been 45 ms ($\mathrm{MBU} = 44\%$, $1.42\times$ the \emph{pessimistic} floor), it would sit outside the interval entirely: no overlap explanation is arithmetically possible, and the tool escalates to Nsight Systems with the three-question reading list, because something outside the resource account (host gaps, stragglers, preemption) is consuming time.

For prefill, the identity of \S\ref{sec:metrics} selects MFU: an 8K prompt costs $2 \cdot 37\mathrm{B} \cdot 8192 \approx 606$ TFLOP of parameter GEMMs, so 16 H20s at 50\% MFU bound TTFT at 256 ms---a GEMM-only lower bound (the causal-attention term adds up to $\sim$25 ms at this length); H100s would bound the same GEMM account at $\sim$38 ms, the prefill face of the ridge portrait.
A measured 400 ms implies 32\% MFU, within the MoE prefill band (\S\ref{sec:thresholds}); the timeline-level suspects (all-to-all exposure, expert imbalance) are checked before any kernel is blamed.

\section{Related Work}

\paragraph{Analytical models and limit studies.}
Pope et al.~\citep{pope2022efficiently} founded the genre: closed-form partitioning selection for dense Transformers on TPUs.
GenZ~\citep{bambhaniya2024genz} maps use cases to platform requirements with a resource decomposition close to ours; LLMCompass~\citep{zhang2023llmcompass} and Calculon-style co-design models~\citep{kundu2024distributedmodeling} serve hardware DSE; roofline treatments cover the single-device~\citep{yuan2024llm} and edge~\citep{bi2026rooflinebench,chen2025elib} settings.
Closest to our decode account, LIMINAL~\citep{davies2025liminal} derives single-user decode limits from bandwidth, compute, capacity, and synchronization, and Erdil~\citep{erdil2025inference} derives token-economics frontiers including a $\mathrm{speed} \propto \mathrm{BW}^{1/3}$ scaling; both fix overlap assumptions and target hardware-evolution questions, where \system{} keeps overlap observable via the $[\max,\mathrm{sum}]$ interval and targets deployment triage.
Yun et al.~\citep{yun2025rethinking} analyze MLA/MoE arithmetic intensity at the operator level, concluding MLA core-attention trends compute-bound; this is consistent with our step-level account---operator intensity and step-level byte totals answer different questions, and after DSA the step is weight-bandwidth-dominated regardless of the attention operator's intensity.
AIC++~\citep{wu2026how} explores the same V3.2 deployment space (including attention--FFN disaggregation, which we do not model) via measured kernel databases plus network simulation on B200; \system{} trades its fidelity for closed forms, interpretable wall ordering, and hardware it has no database for---the two are complementary stages of the same pipeline.

\paragraph{Simulators and configuration search.}
Vidur~\citep{agrawal2024vidur} and KernelSight-LM~\citep{yao2026kernelsight} predict serving performance from profiled operators; LLM-Pilot~\citep{lazuka2024llmpilot} learns black-box cost models across GPUs; AIConfigurator~\citep{xu2026aiconfigurator} searches configurations against a measured-kernel database in seconds.
These tools accelerate the grid search; \system's position is that the floor account should shrink the grid \emph{before} any search, look-up, or simulation---and, unlike a kernel database, closed forms transfer to hardware and models nobody has profiled yet.
When scheduler dynamics or caching dominate, simulators are the correct escalation target, exactly as profilers are for kernel questions.

\paragraph{Serving systems and the DP-attention judgment.}
Phase splitting and scheduling~\citep{patel2023splitwise,zhong2024distserve,agrawal2024sarathi,qin2024mooncake}, module-level disaggregation~\citep{zhu2025megascaleinfer}, hybrid MoE parallelism~\citep{zhou2026mixserve}, and model--system co-design~\citep{stepfun2025step3} define the layout space our accounts judge.
For the MLA replication problem specifically, the engineering record states the mechanism~\citep{sglang2024v04,vllm2025dprfc,deepseek2025profiledata} and architecture papers confirm it while responding by redesigning attention or its sharding~\citep{tang2025tpla,zadouri2025hardwareefficient,meng2026gqla}; reverse-engineering analyses quantify official H800 deployments~\citep{fang2025deepseek}.
None derives the capacity-first judgment for unmodified MLA: the Ant deployment states the memory-bound intuition without quantifying where it binds~\citep{lmsys2025h20}, and TP16-attention deployments remain in production (\S\ref{sec:case})---the gap this case study fills.
H20-specific work exists at the kernel level~\citep{dege2025flashmlaetap} and as empirical throughput reports; the ridge-geometry account of \S\ref{sec:ridge} appears to be new.

\paragraph{MoE metrics and sparsity accounting.}
MoE-CAP~\citep{jiang2024moecap} corrects MBU/MFU for per-token activated parameters; our union correction (\S\ref{sec:union}) additionally captures batch-aggregated weight traffic, using a closed form shared with speculative-decoding and expert-sharing analyses~\citep{huang2025moesd,vankov2026xshare}.
Empirical characterizations~\citep{recasens2025mind,arif2026understanding,chen2026memorybound} document bandwidth saturation, capacity traps, and floor-attainment gaps that our workflow is designed to triage---and that its failure modes (\S\ref{sec:limits}) must respect.

\paragraph{Benchmark discipline and agentic optimization.}
Open-versus-closed-loop bias was formalized by Schroeder et al.~\citep{schroeder2006open}; LLM serving evaluation work has begun rebuilding measurement discipline, and \S\ref{sec:benchdiscipline} applies the classical distinction to serving capacity claims.
Agentic optimization systems increasingly wrap profilers and roofline reasoning in LLM agents~\citep{ding2025asap,gai2026kernelpro,sun2026kernelskill,li2026agentskill,liu2026epoch}; \system's skill artifact is the complementary piece---a deterministic floor layer that disciplines when such agents may escalate to expensive tools.

\section{Discussion}

\paragraph{Why not always simulate?}
Because many bad deployments can be ruled out with one page of arithmetic. If TP16 cannot hold the required concurrency at the target context, no scheduler search fixes that. If a benchmark already sits at $1.27\times$ the optimistic floor, kernel work cannot return a large factor.
Simulators and profilers remain important, but they are stage-2 tools for residuals that survive the floor check.

\paragraph{What generalizes.}
The mental model is not specific to this case study.
The five-dimensional account, the two-sided floor, wall ordering, and the ridge-identity metric discipline apply to any autoregressive transformer served on accelerators: hardware enters through five numbers per device (\S\ref{sec:ridge}), the model through per-module declarations (\S\ref{sec:compositional}), and the deployment through per-module sharding.
Nothing in the workflow references H20, MLA, or MoE specifically---the case study was chosen because it exercises the account's hardest current instance (extreme ridge asymmetry, unshardable latent KV, batch-dependent expert traffic), not because it marks a boundary.
What does \emph{not} transfer automatically: calibration constants (per cluster), the module library (per architecture family), and the blind spots below.
Training-side estimation follows the same resource-vector structure but is out of scope here.

\paragraph{When the floor model fails.}
\label{sec:limits}
The account has known blind spots, and stating them is part of the method.
(i)~\emph{Continuous batching}: $B$ varies within a window, so floors should bracket the batch distribution rather than assume a stationary point.
(ii)~\emph{Expert imbalance}: the union expectation assumes uniform routing; hot experts push per-GPU weight traffic above the account (EPLB-style balancing restores it), and the $\mathrm{sum}$ floor loses meaning when stragglers, not resources, set step time.
(iii)~\emph{Open-loop overload}: past saturation, sojourn time is queueing-dominated; floors bound service time only, and queueing-theoretic capacity models take over.
(iv)~\emph{Host-bound regimes}: Python and scheduler overheads are outside the resource vector by construction---the $<40\%$ MBU band plus kernel-gap inspection exists precisely to catch what the account cannot express.
(v)~\emph{Batch-1 attainment}: launch overheads leave measured single-stream latency far above the floor on high-end parts~\citep{chen2026memorybound}; single-stream floors answer feasibility, not prediction.
(vi)~\emph{Calibration drift}: per-shape GEMM efficiency varies widely; the calibration chain must use production shapes or residuals inherit the error.

\paragraph{Scope.}
The artifact currently models pure TP and EP+DP-attention decode plus compute-bound prefill.
The compositional structure of \S\ref{sec:compositional} defines the extension path: a per-module plan interface admits mixed TP$\times$EP and the production hybrid layouts of \S1; new attention or state-space modules enter by declaring their state and shardability; prefill/decode disaggregation, chunked-prefill interference, speculative decoding, and attention--FFN disaggregation~\citep{wu2026how} are account extensions on the same interface.
All are staged behind validation of the core loop---the framework does not change, the module list grows.

\section{Required Measurement Campaign}
\label{sec:experiments-needed}

The workflow's claims are calibratable, and this section is the paper's own stage-2 escalation plan.
(1)~\emph{Calibration chain} on the target cluster---\emph{executed} (\S\ref{sec:artifact}): production-shape GEMM rates, all-reduce bandwidth/latency sweeps, DeepEP all-to-all benchmarks; the HBM point should be re-taken with \texttt{nvbandwidth} in place of the torch fallback.
(2)~\emph{Residual calibration}: an open-loop, steady-state sweep over $\{$TP16, TP16+DSA, EP16+DPA$\}$ $\times$ $B$ $\times$ $S$ ($\ge$12 points), each reconciled against its floor, yielding the residual distribution that grounds (or revises) the $1.3\times$ threshold and the MBU bands.
(3)~\emph{Triage traces}: one near-floor and one high-residual configuration instrumented with Nsight Systems, demonstrating the three-question reading list and the budget-table indictment of a kernel class.
(4)~\emph{Cross-platform replication} of the workflow on a dense model and a second GPU generation, testing that the method---not the case study---is the contribution.
(5)~\emph{Judgment validation}: the TP16-attention versus DP-attention goodput frontiers measured on the same 16$\times$H20 system, closing the loop on \S\ref{sec:singlestream}'s claim that the production disagreement is an operating-point question.

\section{Conclusion}

\system{} reframes LLM serving optimization as residual-driven triage: compute two-sided floors from a five-dimensional resource account, benchmark only the configurations whose walls permit the target, read the $[\max,\mathrm{sum}]$ position as a free overlap diagnostic, and open profilers only when the residual points to time the account cannot explain.
The H20 case study shows the method's reach: a ridge-point portrait that explains why an export-constrained part is a natural decode engine, a wall ordering that shows its compute ceiling is unreachable behind the KV-capacity wall, a capacity-first derivation of the DP-attention judgment together with its single-stream inversion---and, throughout, arithmetic that fits on a page and reproduces from a dependency-free artifact.
The measurement campaign of \S\ref{sec:experiments-needed} is the remaining step from analytical draft to validated systems paper.

\bibliographystyle{plainnat}
\bibliography{refs}

\appendix

\section{Artifact Commands}

The following commands reproduce Table~\ref{tab:h20-case} (add \texttt{--full-experts} for the conservative union bound shown; omit it for the expectation account):

\begin{verbatim}
cd scripts/
python3 floor.py --model deepseek-v3.2 --gpu h20 \
  --nodes 2 --gpus-per-node 8 --parallel tp16 \
  -B 64 -S 8192 --full-experts

python3 floor.py --model deepseek-v3.2 --gpu h20 \
  --nodes 2 --gpus-per-node 8 --parallel tp16 \
  -B 64 -S 8192 --full-experts --dsa

python3 floor.py --model deepseek-v3.2 --gpu h20 \
  --nodes 2 --gpus-per-node 8 --parallel ep16-dpa \
  -B 64 -S 8192 --full-experts
\end{verbatim}

Single-stream bounds (\S\ref{sec:singlestream}) are printed by the same invocations at \texttt{-B 1}; reconciliation examples:

\begin{verbatim}
python3 mbu_mfu.py decode --model deepseek-v3.2 --gpu h20 \
  --parallel tp16 -B 64 -S 8192 --tpot-ms 25 --full-experts

python3 mbu_mfu.py prefill --model deepseek-v3.2 --gpu h20 \
  --ngpus 16 --prompt 8192 --ttft-ms 400
\end{verbatim}

\end{document}